\documentstyle[preprint,aps,psfig]{revtex}   

\begin{document}
\hsize\textwidth\columnwidth\hsize\csname @twocolumnfalse\endcsname
\title{ On the nature of different types of absorbing states}
\author{Miguel A. Mu{\~{n}}oz}
\address{Dipartimento di Fisica,
Universit\'a di Roma ``La Sapienza'',~P.le
A. Moro 2, I-00185 Roma, Italy}
\date{\today }
\maketitle

\begin{abstract}

   We present a comparison of three different types of Langevin
equation exhibiting absorbing states: 
 the Langevin equation defining the
Reggeon field theory, one
 with multiplicative noise, and a third type
in which the noise is complex. Each one is found to describe
 a different 
underlying physical mechanism; in particular, the nature of the 
different 
absorbing states depends on the type of noise considered.
 By studying the stationary
single-site effective potential, we analyze the impossibility 
of finding a reaction-diffusion model in the multiplicative noise
universality class. We also discuss some theoretical questions related to    
the nature of complex noise, as for example, whether it is necessary or not
 to 
consider a complex equation in order 
to describe processes as the annihilation
reaction, $A +A \rightarrow 0$.

\vspace{10pt}
PACS: 
\vspace{10pt}

\end{abstract}

 
  Different systems and models appearing in physics as well
as in other fields can exhibit absorbing states. 
 An absorbing configuration is one in which a system can get
trapped, from which it cannot escape \cite{Ma,Priv,Granada}.
 Therefore an absorbing configuration
is a fluctuation-free
microscopic state.
 Some examples of systems exhibiting absorbing states,
 among many others, are:
chemical reaction-diffusion models of catalysis \cite{catal}, 
models for the spreading of epidemics or forest fires \cite{epid}, 
directed percolation \cite{Kin,Ma},
 the contact procces \cite{Ha,Priv}, 
models 
of branching and annihilating random walks \cite{baw},
damage spreading \cite{damage}, and even
self-organized systems \cite{PMB}.       

As some control parameter is changed, 
many of these systems experience a phase transition from an 
absorbing phase, i.e., a phase in which the absorbing state
is the only stationary state \cite{Granada},
to an active phase, characterized
by a non-vanishing value of the order parameter.  At the critical
point, these systems exhibit universal features.

It was conjectured some time ago by Janssen and Grassberger 
\cite{conjecture} that all the different systems and models 
with an unique 
absorbing state, a single-component order parameter, and no extra 
symmetry or conservation law belong in the same universality 
class as directed percolation (DP), which is considered the
canonical representative of that vast class of models.
 In a field theoretical description this
universality class is represented by the Reggeon Field theory (RFT),
\cite{RFT}
which in terms of a Langevin equation reads \cite{note3}:
\begin{equation}
{\displaystyle{\partial n (x,t) \over \partial t}} =  \nabla^2 n(x,t) +
 a n(x,t)
-b n^2(x,t) +
 \sqrt{n(x,t)} \eta(x,t)  ~~,
\label{rft}
\end{equation}
where  $n(x,t)$ is a density field at position $x$ and time $t$, 
$a$ and $b$ are control parameters,
and $\eta(x,t)$ is a Gaussian
 noise which only non-vanishing 
correlations are:
 $<\eta(x,t) \eta(x',t')>= D \delta(x-x') \delta(t-t')$.
The equation is interpreted in the Ito sense \cite{Ito}. 
The noise term in  eq.(\ref{rft}) is proportional
to the square root of the field, therefore in the absorbing state
$n(x)=0$ the dynamics is completely frozen: both the deterministic 
and the stochastic terms are equal to zero.
Other higher-order terms could be added to the
 deterministic part of eq.  (\ref{rft}) but they can be easily argued to 
be irrelevant in a renormalization group sense.

  The previous conjecture has been confirmed
 in a large number of computer
simulations and series expansion analysis,
 and  DP-universality class has proven
to be extremely robust against
the modification of many details in the microscopic models.
 The conjecture of universality has been 
extended for multicomponent systems \cite{GG}, as 
well as for systems with an infinite number of absorbing states
\cite{Noi}.

   Nevertheless, not all the systems with absorbing states belong
in the universality class of DP. Some other classes different of 
DP have
been identified, all of them showing some essential physical
 differences with DP.

Two relevant examples for what follows are:

\begin{enumerate}

\item Particle systems
in which evolution occurs only
at the interfaces separating occupied (active)
from empty (absorbing)
regions  
belong to the
{\it compact directed percolation} (CDP)
universality class \cite{CDP}.  Examples of this universality 
class are the one-dimensional diffusion-limited
 reactions of the type $A+A  \rightarrow 0$
and $A+A  \rightarrow A$ (in dimensions larger than  one
 these models are not expected to be in the same universality 
class as CDP).
 The {\it pseudoparticles} $A$ can be thought of
as the {\it kinks} separating active from inactive regions, where
the dynamics occurs.
 A field theoretical description for
such class of systems
was proposed by Peliti in \cite{Peliti,DKCA}; its equivalent 
Langevin equation reads:
\begin{equation}
{\displaystyle{\partial n (x,t) \over \partial t}} =  \nabla^2 n(x,t) 
-b n^2(x,t) +
 i n(x,t) \eta(x,t)  ~~,
\label{pft}
\end{equation}
where  $n(x,t)$ is a field,  $i$ is the imaginary unit, and
$\eta$ is a Gaussian noise with some amplitude $D$. Note that 
$n(x)=0$ is an absorbing configuration.
Hereafter we refer to eq. (\ref{pft})  as Peliti's field 
theory.  It is important to point out that the field $n(x,t)$
is not the density of $A$ particles, but a more abstract field
\cite{note2} 
which expectation value coincides with that of the real density field,
and which higher order moments can be also related to higher order
moments of the density field. All this will become clearer in 
a forthcoming section where the explicit derivation of eq. (\ref{pft})
is performed.

  Before concluding this epigraph let us point out that there is a class
of systems which present an active as well as an absorbing phase with
 a phase transition separating both of them, and which associated
 noise should also present a complex structure: this is 
the so called {\it parity-conserving universality class} \cite{baw}.
Due to the extra conservation law it is clear by now that these systems 
are not in the RFT universality class.

\item  A new universality
class characterized
by a noise different from that of the
previously described classes has recently been elucidated
( see \cite{Pik,MN,DNA} and references therein):
 the  {\it multiplicative noise}
(MN) universality class.
While in the Langevin equation for the RFT
 the noise amplitude is proportional to the square 
root of the field at each point,
 in the MN universality class
the noise amplitude is proportional to the field itself, i.e.
\begin{equation}
{\displaystyle{\partial n (x,t) \over \partial t}} =  \nabla^2 n(x,t) +
 a n(x,t)
-b n^2(x,t) +
 n(x,t) \eta(x,t)  ~~,
\label{mn}
\end{equation}
with $n(x,t)$ being a density field at position $x$ and time $t$,
and $\eta(x,t)$ a Gaussian white noise. Obviously this new type
of noise is also compatible with the presence of an absorbing state
at $n(x)=0$ at which the dynamics is completely suppressed.
\end{enumerate}

  While there are many microscopic reaction-diffusion models
belonging in the DP universality class, 
and is also easy to identify
reaction-diffusion models in the Peliti's
field theory, 
no microscopic reaction-diffusion model in the MN universality class
has been identified so far.
 The possibility of constructing a reaction-diffusion model that 
exhibits the rather striking properties of MN \cite{MN}
has been explored in a recent 
paper by Howard and T\"auber \cite{HT}. They concluded that 
given the apparent impossibility of finding such type of model
 in the MN class, the physical meaning
 of that universality class is
unclear.
 
Motivated by the previous work \cite{HT}
 we have further investigate
this issue. In what follows we present a comparison
of the different  noise terms
 appearing in Langevin equations for Reggeon
field theory, the multiplicative noise, and Peliti's field theory,
to identify physical differences among them.
A simple and intuitive justification of the fact that 
no reaction-diffusion system can be found in the MN universality
class is given.
 Alternatively, we enumerate some other discrete, microscopic
 models belonging to that class. We also discuss some curious 
properties of systems with complex noise, and analyze whether a real
Langevin equation can be written for systems like the annihilation
reaction $A+A \rightarrow 0$.

\section{Analysis of the single-site effective potential}

\subsection{Reggeon field theory}

\begin{figure}
\centerline{\psfig{figure=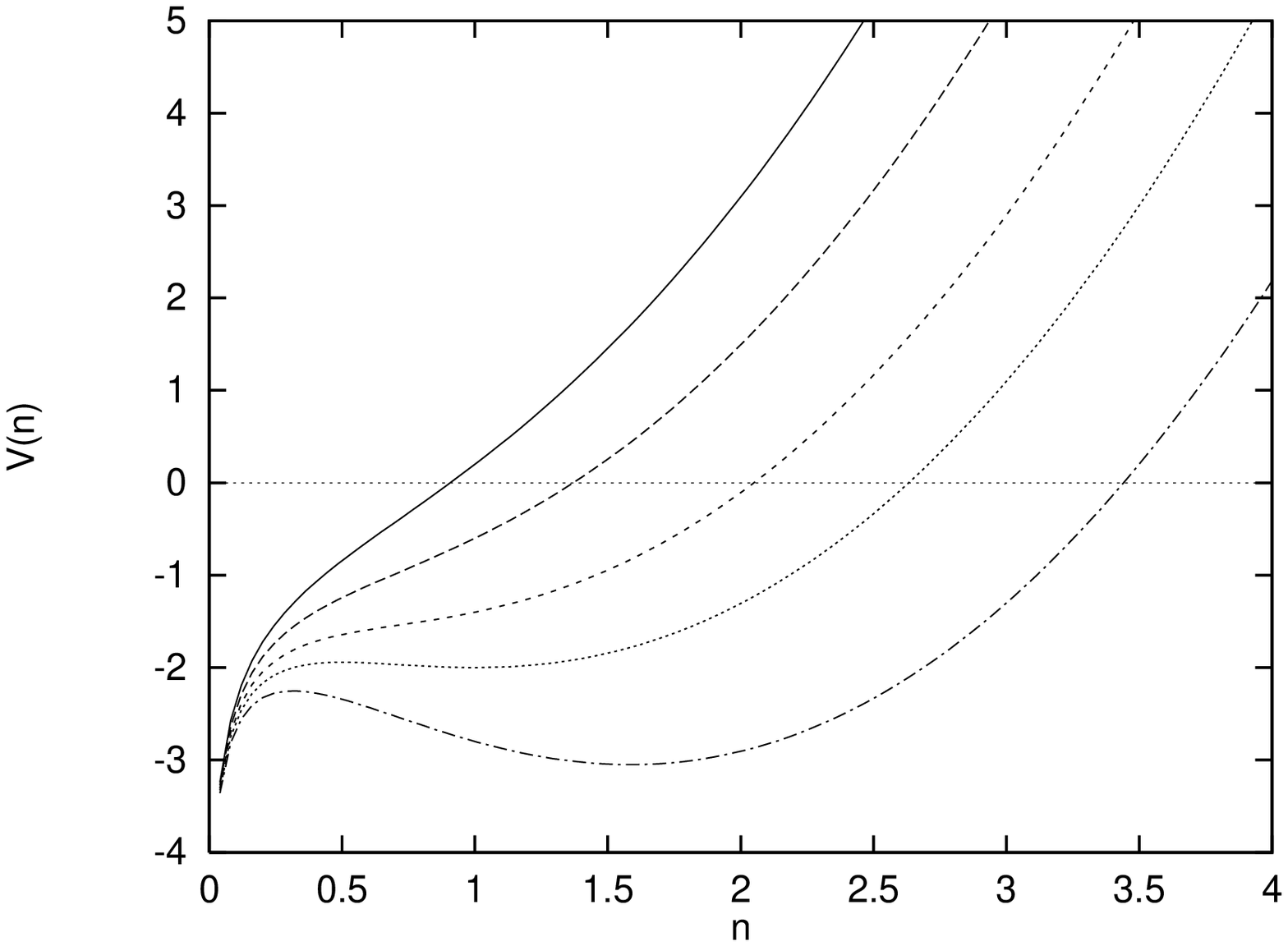,width=11cm}}
\caption{ \label{V0rft}  Potential $V(n)$ associated to the 
formal stationary 
solution of the 0-dimensional RFT  for different parameter values:
$b=D=1$, and from top to bottom: $a=0.4, 0.8, 1, 1.5$ and $1.9$.
Note the presence of a strong (non-integrable)
singularity at the origin in any case.}
\end{figure}

 We start by analizing the zero-dimensional (single variable)
 version of 
the RFT Langevin equation.
  The Fokker-Planck equation associated to eq. (\ref{rft}) is 
\cite{Ito}:
\begin{equation}
{\displaystyle{\partial P (n,t) \over \partial t} }= 
-{\displaystyle{\partial \over \partial n} } (a n -b n^2) P(n,t)
+ {\displaystyle{D \over 2} }
{\displaystyle{\partial^2 \over \partial n^2} } n P(n,t).
\label{fprft}
\end{equation}
 By imposing the detailed balance condition, the 
associated formal stationary probability
 distribution is found to be :
\begin{equation}
P(n) = \exp (-V(n)) \propto {\displaystyle{1 \over n} } 
\exp [{\displaystyle{ 2 \over D } }(a n - b n^2/2)] 
\label{Prft0}
\end{equation}
where $V(n)$ is the effective potential. 
In figure \ref{V0rft} we plot $V(n)$ for different 
values of $a$, and $D=1$;  
for $ a \ge \sqrt 2 b $ the potential has a minimum 
at $n \neq 0$, while for $a < \sqrt 2 b$ $V(n)$ has no maximum 
or minimum.
Note, however, that due to the (non-integrable)
 singularity at $n=0$,
the probability  eq. (\ref{Prft0}) is not normalizable, and the
only stationary solution is $P(n)= \delta (n)$. Therefore
there is no active phase in this simple 0-dimensional case,
and the systems decays towards the absorbing state for any
set of paramenter values.

   Let us now study how the single-site effective potential 
behaves in dimensions larger than zero.
 In particular, we perform a numerical     
simulation of eq. (\ref{rft}) in one dimension. To do so we employ a
technique developed by Dickman \cite{discre} to deal with
numerical simulations of the continuous RFT.
Let us point out that the simulation of this continuous theory
with an absorbing state is not a trivial issue, and that, for
example, a straightforward
discretization of eq. (\ref{Prft0}) in which eventual negative 
values of the field  (that may appear due to the discretization)
are fixed to $n(x)=0$,  does not preserve the presence of an
absorbing state.
Dickman's method consists of a
 discretization of the space, time and also of the field
variable, that ensures the presence of an absorbing state
(see \cite{discre} for details).

\begin{figure}
\centerline{\psfig{figure=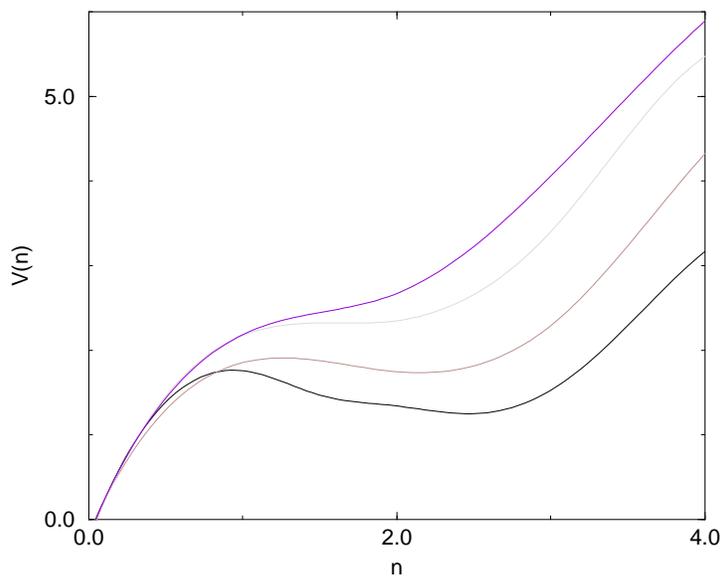,width=11cm}}
\caption{ \label{V1rft}  Stationary potential for the
one-dimensional RFT as coming from a
simulation of the discretized Langevin equation.
The uppermost curve corresponds to a value
of a in the absorbing phase; the second one to $a=a_{critical}$
while the two lower ones are in the active phase.
}
\end{figure}

In the one-dimensional case 
 the active phase survives to the effect of fluctuations contrarily
to what happens in the single-variable case.
In figure \ref{V1rft} we show the effective potential (defined
as minus the logarithm of the normalized stationary probability 
distribution) for 
different values of $a$. The uppermost curve corresponds to a value
of a in the absorbing phase; the second one to $a=a_{critical}$
while the two lower ones are in the active phase.  
Note that in all the cases a singularity at the origin of the same type
 is present, and consequently,
for any finite system there is a finite probability for the 
system in the active phase to go through the potential barrier and
decay towards the absorbing state: the active phase is a metastable
state.  The mean 
time required for the system to overpass the
barrier and collapse to the absorbing 
state grows exponentially with time, and becomes infinite in 
the thermodynamic limit. In this way the phase transition appears
 only in infinitely large systems, and the large system-size 
limit has to be taken first than the infinite time limit in
order to permit the presence of an active phase.
In dimensions larger than $d=1$ the same qualitative type of
behavior is expected.

\subsection{Multiplicative noise}

\begin{figure}
\centerline{\psfig{figure=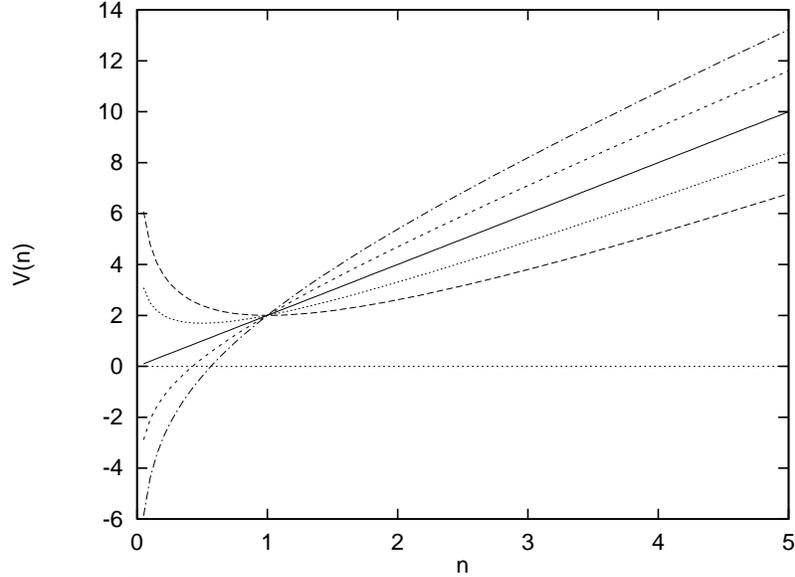,width=11cm}}
\caption{ \label{V0mn}  Potential $V(n)$ associated to the
 formal stationary
solution of the 0-dimensional multiplicative noise equation
for different parameter values:
$b=D=1$, and from top to bottom: $a=2, 1.5, 1, 0.5$ and $0$.
The potential develops a minimum as $a$ is increased, has
a negative singularity at the origin for $a<1$, and a positive
singularity for $a>1$; the singularity is integrable in the
active phase, i.e., when $a<0.5$, while in the absorbing phase
the only stationary solution is a delta function at the origin.
}
\end{figure}

  The Fokker-Planck equation associated to eq. (\ref{mn}) in
the zero-dimensional case  is:
\begin{equation}
{\displaystyle{\partial P (n,t) \over \partial t} }=
-{\displaystyle{\partial \over \partial n} } (a n -b n^2) P(n,t)
+ {\displaystyle{D \over 2} }
{\displaystyle{\partial^2 \over \partial n^2} } n^2 P(n,t).
\label{fpmn}
\end{equation}
 By imposing the detailed balance condition, the
stationary formal solution is:
\begin{equation}
P(n) = \exp (-V(n)) \propto {\displaystyle{1 \over n^{2(D-a)/D}} }
\exp [{\displaystyle{-2 b n \over D} }]
\label{Pmn0}
\end{equation}
where $V(n)$ is the effective potential; the solution is not 
normalizable when $a < D/2$, and normalizable otherwise.
$V(n)$ is plotted in figure \ref{V0mn} for different
values of $a$ and $D=1$;
 the two lowermost curves correspond to eq. (\ref{Pmn0})
in the absorbing phase ($a=0$ and $a=0.5$)
(where the only stationary solution is 
$P(n)=\delta(n)$). The central one ($a=1$), and
the two uppermost curves ($a=1.5$ and $a=2$)
 are in the active phase. Note that 
contrarily to the RFT the MN exhibits a phase transition even
in zero dimensions. Observe also that the singularity at the origin
in the formal solution  eq. (\ref{Pmn0}) changes its degree
as $a$ is increased, 
in contrast with what happens for
eq. (\ref{Prft0}); in fact, for $D/2<a<D$ (in the active phase)  the
 singularity is integrable, and 
above $a=D$ (also in the active phase)
 the origin becomes {\it repelling} instead
of {\it absorbing}. This is an essential difference with RFT.

\begin{figure}
\centerline{\psfig{figure=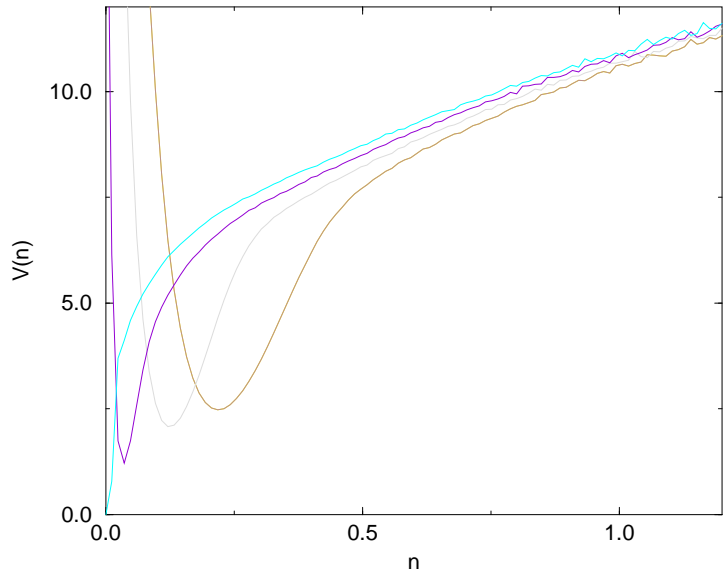,width=11cm}}
\caption{ \label{V1mn}  Stationary potential for the
one-dimensional multiplicative noise Langevin equation
as coming from a
simulation of the discretized equation. The curves       
correspond to four different parameter values all of them 
in the active phase. 
Note that the singularity at the origin 
is positive for the three lower most curves ($a=-2$, $-2.1$ and $-2.2$
respectively),
 therefore there is not an absorbing, but a repelling
state. 
The 
uppermost curve ($a=-2.23$),  with a negative singularity at the origin,
is still in the active phase, but the singularity is integrable.
}
\end{figure}

Let us now explore how this property of the single-site potential 
is modified in higher dimensions. For that, we perform a 
numerical integration of the stochastic equation defining the
model, which presents
less technical difficulties than the integration of the RFT \cite{MN}.
 The result are presented in figure \ref{V1mn}. 
Qualitatively  the potential shape changes in the same way as it does 
 in the 0-dimensional
case.
 Above the critical point, there is either an integrable
singularity (uppermost curve) or a repelling wall (three other curves) 
at the origin.
 The first case, i.e., an integrable singularity
at the origin occurs in a very tiny region of the parameter space.
On the other hand, 
in the absorbing phase there is a collapse of the probability 
towards $P(n)=\delta(n)$ (non-integrable singularity at the origin).

 Therefore, the physics is very 
different from that of RFT: 
when the system is in the active phase, there is either 
an integrable singularity at the origin of the potential or
a repelling wall. In any case, the situation differs
from that in RFT, in which there are two locally stable attractors
in the active phase: one at the origin (the stable one) and one
at a different point (a metastable one).

   That is the reason  why a reaction-diffusion system can not
be described by an equation as
 eq. (\ref{mn}): in a finite reaction-diffusion
system with a non-vanishing particle annihilation rate there is 
 a non-zero probability of reaching the absorbing (empty)
state for any finite system-size and for any set of parameter values.
 In systems with MN there is 
no accessible absorbing state in the active phase, i.e. there 
is no non-integrable singularity at the origin of the potential 
(and consequently no collapse of the probability density to the origin),
 and therefore
MN does not capture the physics of reaction-diffusion systems.
  
That fact, does not mean that it is not possible to construct
discrete lattice model in the multiplicative noise universality class.
In particular, systems exhibiting an unbinding transition from a
wall (as, for example, the problem of local alignment
 of DNA chains \cite{DNA}, and wetting transitions \cite{Livi})
 belong in this universality class. These models are
 usually defined in terms of a field 
variable $ h(x,t)= \pm \log(n)$ that flows to $\mp \infty (n=0)$ in the 
absorbing phase, and that reach a non-vanishing stationary
 average value 
otherwise.

 \subsection{Peliti's field theory}

  In this section we compare the effect of the complex
  noise appearing in the Peliti's field theory  eq. (\ref{pft}), with
the two previously studied cases
to get a global picture of the
different type of noises that can appear in systems with 
absorbing states. In order 
to understand what is the origin of the 
complex noise in processes as $A+A \rightarrow 0$,
 and to clarify whether a microscopic
 process like that with a real density field
has necessarily to be described by a complex Langevin equation,
 we present a derivation of the Peliti's field
theory for the annihilation reaction: $A+A \rightarrow 0$ by employing
the exact Poisson representation introduced by Gardiner and Chatuverdi
\cite{Gardiner,note2}. Note that for this reaction there is no active
phase, and the interesting magnitudes are those describing the
decay towards the absorbing state.
  For the sake of simplicity
in the notation we present here the 0-dimensional case,  extensions
to higher dimensions being straightforward.

   The master equation defining the process is:

 \begin{equation}
{\displaystyle{\partial P (n,t) \over \partial t} }=
k[(n+2)(n+1) P(n+2,t)-n(n-1) P(n,t)]
\label{master}
\end{equation}

  Multiplying both sides  of  eq. (\ref{master}) by $s^n$, summing
over all $n's$ from $0$ to $\infty$, and defining the generating
function $G(s,t)= \sum_{n=0}^{\infty} s^n P(n,t)$, we get
\begin{equation}
{\displaystyle{\partial G (s,t) \over \partial t} }=
k (1-s^2) \partial_s^2 G(s,t).
\label{master2}
\end{equation}
We now introduce the Poisson transformation:
\begin{equation}
P(n,t)= \int d \alpha  {\displaystyle { \alpha^n \exp(-\alpha) 
\over n!} } f(\alpha,t)
\end{equation}
where $f(\alpha,t)$ is a given function (see \cite{Gardiner}),
in terms of which : 
\begin{equation}
G(s,t) = \int d \alpha f(\alpha,t) exp(\alpha (s-1)).
\end{equation}

\begin{figure}
\centerline{\psfig{figure=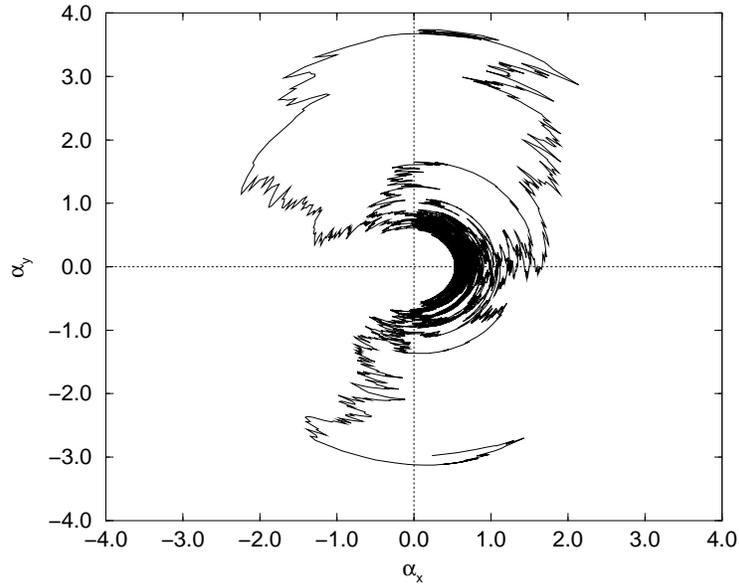,width=11cm}}
\caption{ \label{trajectory} Typical trajectory of the complex
noise equation in the stationary regime. 
 Note that there is a large probability  of finding
the system in the vicinity of $1/2,0$, while points inside a circle
centered at the origin of radius $1/2$ are inaccessible. 
}
\end{figure}

\begin{figure}
\centerline{\psfig{figure=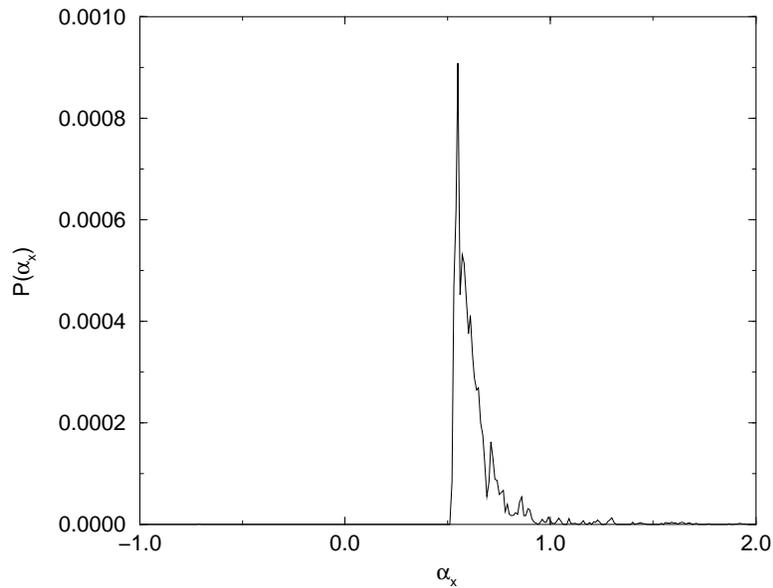,width=11cm}}
\caption{ \label{corte} Cut of the stationary probability
distribution associated to the complex noise equation
 with $\alpha_y=0$.
Observe the non-differentiability at $\alpha_x=1/2$.
}
\end{figure}

The Poisson transformation has the interesting
 property that the moments of $P(n)$ and $f(\alpha)$
 can be easily related:
$<\alpha^p>=<n(n-1)...(n-p+1)>$; in particular the first moments
are the same for both distributions.
 The integral over $\alpha$ can be taken over different domains of 
integration; for
the moment let us assume $\alpha$ to be a real variable and
 leave the integration domain undetermined.
In terms of $f(\alpha,t)$,  eq. (\ref{master2}) reads, 
\begin{equation}
\int d \alpha \exp( \alpha (s-1) ) \partial_t f(\alpha) =
- k \int d \alpha  \alpha^2 f(\alpha,t) [-\partial_\alpha^2 +2 
\partial_\alpha] \exp( \alpha (s-1) )
\end{equation}
which, integrating by parts, and assuming that the boundary terms give 
a vanishing contribution to the integral 
 \cite{note1} can be written as:
\begin{equation}
\partial_t f(\alpha) =
 k [ 2 \partial_\alpha  \alpha^2 f(\alpha,t)
    -\partial_\alpha^2 \alpha^2 f(\alpha,t)]
\label{fpp}
\end{equation}
which is a Fokker-Planck equation with a negative diffusion coefficient.
The Langevin equation stochastically equivalent to the previous
Fokker-Planck equation is: 
\begin{equation}
 \partial_t {\alpha}(t) = 2 \alpha(t)^2 + i  \sqrt 2 \alpha \eta(t)
\label{lan}
\end{equation}
where $\eta(t)$ a Gaussian noise with amplitude $1$,  $i$ 
the complex unit, and $k$ has been eliminated by redefining
the time as $t \rightarrow  t/ k$.
Note that we have arrived to an inconsistency: 
$\alpha $ was assumed to be a real variable
 and we have arrived to a complex equation
(observe that due to the complex
 term in eq. (\ref{lan}) $\alpha$ develops
an imaginary part even if it is taken to be real at time $t=0$.
In appendix A we present further details on the impossibility
of defining a real Poissonian representation for the reaction
$A+A \rightarrow 0$).
Let us now repeat the previous program but performing a complex
transformation instead of a real one, i.e., we take $\int d \alpha $ to  be
 $\int_{-\infty}^{\infty}
  d \alpha_x  \int_{-\infty}^{\infty}  d \alpha_y$,
where $\alpha_x$ and $\alpha_y$ are the real and imaginary parts of
$\alpha$ respectively. This type of transformation leads 
to a function $f(\alpha)$ which is positive and can be identified 
as a probability distribution \cite{Gardiner}.
Proceeding in that way we get a new set 
of Langevin equations for
the variables $\alpha_x$ and $\alpha_y$ 
\begin{eqnarray}
\partial_t {\alpha_x} (t) & =&
 -2  (\alpha_x^2 (t) - \alpha_y^2 (t))  + \sqrt 2 \alpha_y(t) \eta(t)    
\nonumber \\
\partial_t {\alpha_y} (t) & =& -4 \alpha_x(t) \alpha_y(t) - 
\sqrt 2 \alpha_x(t) \eta(t)
\label{complex}
\end{eqnarray}
that is equivalent to the original master equation.
Note that both of the equations in eq. (\ref{complex})
 include the same noise function $\eta(t)$,
and that they could be obtained straightforwardly
 from eq. (\ref{lan}) just by writing $\alpha=\alpha_x + i \alpha_y$
and separating the real and imaginary parts). 
A typical trajectory of the previous set of equations in the
stationary state is shown 
in figure \ref{trajectory}; it wonders in the complex plane
avoiding a region around $0$. Even if the stationary solution
of the underlying process $A+A \rightarrow 0$ is a delta 
function at zero or one particles (depending on whether the
initial condition is even or odd respectively), 
the stationary probability associated to eq.(\ref{complex}) 
is not a delta 
function, but some complicate distribution with $<\alpha_x> = 1/2$ and 
$<\alpha_y> = 0$. 
 The value $1/2$ comes from the fact that for initial
conditions with $n$ even, $n(t \rightarrow \infty) = 0$ and, for
$n$ odd, $n(t \rightarrow \infty) = 1$; the variable
 $<\alpha_x> =<n>$ is the
average of the two previous possibilities.
 On the other hand, the expectation
value of the imaginary part is zero as expected given the relation
among moments of $f(\alpha)$ and $P(n)$.
 It is interesting to note that the
effective potential associated to the stationary distribution 
is a non-differentiable one; in figure (\ref{corte})
we show a one-dimensional cut of the stationary
potential for different $\alpha_x$ with $\alpha_y=0$
 as computed in a simulation of eq. (\ref{complex}).     
Note that contrarily to the cases of the RFT and multiplicative
noise equation, now the dynamics is not frozen
 even if the system has relaxed to the absorbing state.

 We now explore the possibility of finding a real-variable 
 Langevin equation describing this class of systems \cite{note4}.

\begin{figure}
\centerline{\psfig{figure=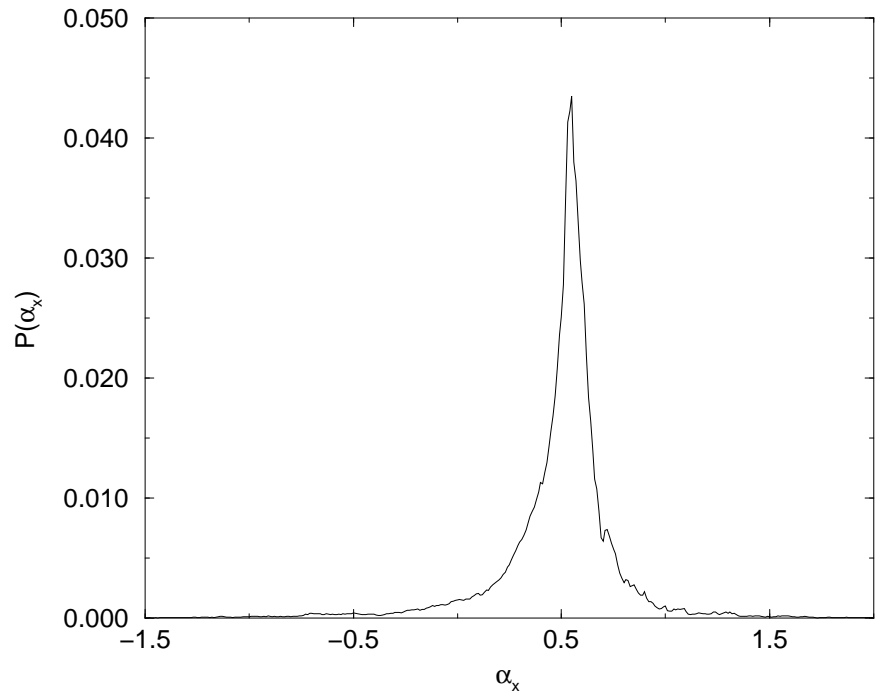,width=13cm}}
\caption{ \label{histoProyect} Stationary probability 
distribution function associated to the non-Markovian equation
or equivalently, the projection of the stationary probability 
function associated to the complex noise equation
 over the real axis.
}
\end{figure}

  As the second equation in eq. (\ref{complex}) is linear in $\alpha_y$
it is possible to integrate it analytically; doing so and substituting 
the result in the 
first one, we get a closed equation for $\alpha_x$, that reads:
  \begin{equation}
\partial_t {\alpha_x} (t)  = \alpha_x 
 -2  \alpha_x^2 (t)+ 2 I(t)^2 + \sqrt 2 I(t) \eta(t) 
\label{nonMark}
\end{equation}
with
\begin{equation}
I(t)= \alpha_y(0) 
  \exp( - \int_{0}^t dt' ( 4\alpha_x -1)) 
-\sqrt 2
 \int dt' \alpha_x(t') \eta(t')
 \exp( - \int_{t'}^t dt'' ( 4\alpha_x -1))
\label{I}
\end{equation}
which is a non-Markovian equation (see appendix B).
The stationary potential associated to eq.(\ref{nonMark}) cannot
be calculated analytically; the numerical solution is shown in figure
\ref{histoProyect}. First we observe that it is 
non-differentiable at $\alpha_x=1/2$; also  we point out that
$\alpha_x$ is not absorbing in general (except for the pathological 
and unphysical case
$\alpha_y(0)=0$), in other words:
 due to the presence of the non-Markovian terms, proportional to
 $I(t)$,
the system can cross from positive values to negative ones.

The role of the complex variable
in eq. (\ref{lan})
is played, after $\alpha_y$ has been integrated out,
 by the non-Markovian
terms in eq. (\ref{nonMark}), and in both cases the absorbing state
of the microscopic associated process is not described by a frozen
dynamics in the Langevin representation, but by a non-trivial 
dynamics (complex or non-Markovian) which statistical properties
reproduce those of the reaction-diffusion model.
   Therefore, the nature of the absorbing state in this case 
is essentially different from those of the previously studied cases.

 For the sake of completeness let us just mention briefly that
a numerical study of the one-dimensional
 Peliti's field theory in terms of
a complex Langevin equation has been recently published 
\cite{Tomeu}. The measured magnitudes are in very good agreement
with the theoretical predictions coming from renormalization-group
and other type of analysis \cite{Peliti,Lee},
 confirming that a complex representation
captures the physics of microscopic systems as, for example, the 
procces
$A+A \rightarrow 0$.


\section{Conclusions}
  We have analyzed different Langevin equations associated to 
systems with absorbing states.
Systems described by the Reggeon field theory Langevin equation
exhibit a non-integrable singularity at the origin
 of the single-site potential,
that corresponds to a true absorbing state, i.e., there
is an accumulation of probability density at the origin,
 while the active state
is a metastable one for finite system-sizes.
 Systems with multiplicative
noise instead change the degree of the singularity at the origin
as the control parameter is changed:
while in the absorbing phase there is a collapse of the
probability density towards the origin, in the active phase there
is either an integrable singularity at the origin or it 
becomes repelling, in which case the 
 the probability to be nearby the origin 
becomes
extremely small, and there is no accessible absorbing state.  
  That is the reason why is not possible 
to find reaction-diffusion systems (in which for finite size systems 
there is always a finite probability of reaching the absorbing state)
 in the multiplicative noise
universality class.
We have also analized some aspects of the annihilation process $A+A
\rightarrow 0$, which is described by a complex noise Langevin equation
or alternatively by a real non-Markovian Langevin equation. 
This type of Langevin equation shows a behavior quite different from
that of Reggeon field theory and multiplicative noise; in particular,
even if the system is in the absorbing state, there
is not a collapse of the probability density to a delta function, and
the dynamics is non-trivial. Systems with complex noise can 
alternatively be described by real non-Markovian equations.

\section{Appendix A}
 
Let us consider the pair of reactions $A+A \rightarrow 0$
and $A \rightarrow 2A$, the first occurring with a rate $k_2$
and the second with $k_1$. The associated master equation is 
\begin{eqnarray}
{\displaystyle{\partial P (n,t) \over \partial t} } &=&
k_2[(n+2)(n+1) P(n+2,t)-n(n-1) P(n,t)] \nonumber \\
 &+ & k_1[(n-1) P(n-1)- n P(n)].
\label{master22}
\end{eqnarray}
Performing a real Poissonian transformation we get:
\begin{equation}
\partial_t f(\alpha) =
  [  \partial_\alpha   (k_1 \alpha - 2 k_2 \alpha^2) f(\alpha,t)
    +\partial_\alpha^2 (k_1 \alpha -   k_2 \alpha^2) f(\alpha,t)]
\label{fp2}
\end{equation}
which is equivalent to the Langevin equation 
\begin{equation}
 \partial_t {\alpha}(t) = (k_1 \alpha - 2 k_2 \alpha^2) + 
  \sqrt 2 (k_1 \alpha -   k_2 \alpha^2)^{1/2} \eta(t)
\label{lan2}
\end{equation}
interpreted in the Ito sense.
Note that the factor multiplying the noise is positive in the
interval $\alpha \in ]0,k_1/k_2[$,
 and vanishes at the limits of the previous
interval.  The formal stationary solution of eq. (\ref{fp2}) is :
 \begin{equation}
  f(\alpha) \propto  { \displaystyle 
1 \over \alpha } \exp(2 \alpha)
 (1-\alpha \beta)^{(1-\beta)/\beta}
\end{equation}
 with $\beta=k_2/k_1$. 
 Considering the Poisson representation as defined 
in $[0, k_1/k_2]$, it is a matter of simple algebra to verify that
the boundary terms appearing in the processes of getting eq. (\ref{fp2})
from eq. (\ref{master22}) give a vanishing contribution. 
At the same time, trajectories of eq. (\ref{lan2}) with initial
condition in $[0, k_1/k_2]$ do not leave that interval. 
On the other hand, if the domain of integration was extended
over those limits, eq. (\ref{lan2}) would develop an imaginary part
and the procedure would not be self-consistent.
 Therefore
the transformation is
 well defined only in the real interval $[0, k_1/k_2]$. 
From a renormalization group point of view the noise-term proportional
to $k_2$ in eq. (\ref{lan2}) can be argued to be irrelevant
rendering the system in the RFT universality class.

 We can now take the limit $k_1 \rightarrow 0$ to see what happens
in the Peliti's field theory case: the interval in which the
Poisson representation is defined shrinks down to a single point;
 $\alpha=0$.  
In the strict 
limit $k_1=0$, a meaningful
 real Poisson representation can not be performed,
and a complex representation is required.

  Let us point out as a final remark that it is somehow surprising
that the standard renormalization group analysis of the Peliti's
field theory, based on a path integral representation of eq. (\ref{lan})
(or equivalently of eq.(\ref{fpp})), in which $\alpha$ is treated
 as a real 
variable, give the right exponents and properties \cite{Peliti,Lee}.
We will investigate that apparent paradox in a future work.

\section{Appendix B}

As a last attempt to write down a one-variable Langevin
 equation with a structure simpler than eq. (\ref{nonMark}),
 and inspired by the rotational quasi-symmetry of the stationary
 distribution solution (see figure \ref{trajectory}), we
perform a change of variables to polar coordinates $\rho$ and $\theta$
defined by:
$ \alpha_x = \rho \cos( \theta)$, $ \alpha_y = \rho \sin( \theta)$.
After changing variables (for which Ito calculus  is required 
\cite{Ito}), we get:
\begin{eqnarray}
\partial_t \rho (t) &=& \rho(t) - 2 \rho^2(t)
 \cos( \theta) \nonumber \\
\partial_t \theta(t) &=& -2 \rho(t)
\sin(\theta) - \sqrt 2 \eta(t).
\end{eqnarray}
Observe that the first one is a deterministic equation, while the second
one is stochastic.  It is easy to verify that this system does not
admit a potential solution (which is consistent with the stationary
potential being non-differentiable). This new set of equations
permits to derive analytically some of the properties of the
stationary probability distribution (as, for example, the presence of
a maximum at $(1/2,0)$), but it does not simplify
the elimination of one of the variables
 in favor of the other one to construct a simple
 one-variable Langevin equation.


\section{Acknowledgements}

 It is a pleasure to acknowledge  Yuhai Tu,
Geoffrey Grinstein, Ron Dickman, Pedro Garrido and Andrea Gabrielli
for very useful discussions and comments. I am grateful to the    
anonymous referee whose comments and criticisms helped me
to improve this paper. 
 This work was
 supported by the European Community through 
grant ERBFMBICT960925.

\newpage

\newpage
 
\newpage

\newpage

\begin{references}
\bibitem{note3} An equivalent, and probably less confusing,
 way to write down the Langevin equation
for the RFT
consists of omiting the square-root of $n$ term in it and substituting 
$D$ by $D n(x,t)$ in the noise correlator.
\bibitem{note4} Langevin equations describing the annihilation process 
in terms of real variables have already appeared in the literature; 
see \cite{conjecture} and references therein.
\end{references}
\end{document}